\documentstyle[12pt,epsfig]{article}
\makeatletter
\def\section{\@startsection {section}{1}{\z@}{-1cm plus-1ex
    minus-.2ex}{1.5ex plus.2ex}{\reset@font\normalsize\bf}}
\def\subsection{\@startsection{subsection}{2}{\z@}{-0.7cm plus-1ex
    minus-.2ex}{1.0ex plus.2ex}{\reset@font\normalsize\it}}
\def\subsubsection{\@startsection{subsubsection}{3}{\z@}{-0.7cm plus-1ex 
    minus-.2ex}{1.0ex plus.2ex}{\reset@font\normalsize}}
\def\paragraph{\@startsection
     {paragraph}{4}{\z@}{3.25ex plus1ex minus.2ex}{-1em}{\reset@font
     \normalsize\bf}}
\def\subparagraph{\@startsection
     {subparagraph}{4}{\parindent}{3.25ex plus1ex minus
     .2ex}{-1em}{\reset@font\normalsize\bf}}

\ifcase \@ptsize
\or % mods for 11 pt 
\or % mods for 12 pt
\fi
\advance\textheight by \topskip
\advance\footskip by 6mm

\newcount\dfttsuboption 
\def\dfttnum#1{\def\@dfttnum{#1}}
\dfttnum{??/xx}
{ \newcount\hour \newcount\minutes \newcount\hourint%
  \minutes=\time \hour=\time \divide\hour by 60%
  \hourint=\hour   \multiply\hourint by -60  \advance\minutes by \hourint%
  \xdef\@time{\the\hour:\ifnum\minutes>9\else0\fi\the\minutes}%
}

\def\@maketitle{\newpage
 \null\begingroup\def\baselinestretch{1}
 \ifcase\dfttsuboption  %=0, standard preprint
   {\raggedleft\normalsize DFTT \@dfttnum \par}%
 \or %=1, draft version of paper
   {\centering\normalsize \draftname 
       \hfill\@date~---~\@time\hfill DFTT \@dfttnum\par}%
 \or %=2, internal memo
   {\centering\normalsize \draftname \hfill \@date\par}%
 \fi
 \vskip 0.8cm
 \begin{center}%
  {\Large \@title \par}%
  \vskip 1.5em
  {\normalsize
   \lineskip .5em
   \begin{tabular}[t]{c}\@author
   \end{tabular}\par}%
  \ifnum\dfttsuboption=0 \vskip 1em{\footnotesize \@date}\fi%
 \end{center}%
 \par\endgroup
 \vskip 1cm}

\newenvironment{summary}{\begin{quote}\begin{center}\bf\abstractname\par%
\end{center}\vskip 0.25em}{\end{quote}\vskip 2em}

\renewcommand{\[}{\begin{equation}}
\renewcommand{\]}{\end{equation}}

\def\eref#1{(\ref{#1})}
\makeatother

\dfttnum{46/94}
\newcommand{\E}{\mathrm{e}}         % Euler's number
\newcommand{\ee}{$e^+e^-$}          % e+ e- (annihilations...)
\newcommand{\nbar}{\bar n}        % NB parameters: must be used in math mode
\newcommand{\NF}{\cal{N}_{\kern -1.9pt f}}     %math mode only
\newcommand{\NC}{\cal{N}_{\kern -1.7pt c}}     %math mode only
\newcommand{\pT}{{p\kern -.2pt\lower 4pt\mbox{\tiny T}}}    %works?
\newcommand{\pL}{{p\kern -.2pt\lower 4pt\hbox{\tiny L}}}    %works?
\title{ \bf Properties of factorial cumulant to factorial moment ratio
  \thanks{\it Work supported in part by M.U.R.S.T. (Italy) under grant 1993.} }
\author{R. UGOCCIONI, A. GIOVANNINI and S. LUPIA \\
 \vspace{0.2\baselineskip}
 \it Dipartimento di Fisica Teorica and I.N.F.N. - Sezione di Torino\\
 \it via P. Giuria 1, 10125 Torino, Italy}
\date{ October 11, 1994 }

\begin{document}

\maketitle

\begin{summary}
It is shown that the ratio of factorial cumulant moments to factorial
moments for a multiplicity distribution truncated in the tail reveals 
oscillations in sign similar to those observed in experimental data. 
It is suggested that this effect
be taken into account in the analysis of data in order to obtain
correct physical information on the multiplicity distributions.
\end{summary}

\section{Introduction}

The study of multiplicity distributions (MD's) has revealed an important
empirical regularity, as they are well described by a negative binomial
distribution (NBD), 
in all reactions\cite{Schmitz} (from \ee\ annihilation to nucleus-nucleus
collisions), in full phase space (at the highest
energies, after disentangling events with different jet topology in \ee\
annihilation\cite{Del:4} and
different components in $\bar pp$ collisions\cite{Fug} ) 
and in symmetric rapidity intervals.
On one hand, the fact that the NBD fits all available data
within 10\% should be emphasized\cite{Singapore}; 
on the other hand, deviations from the fits should be carefully
studied. One possible deviation lies in the exact shape of the tail of
the distribution, which can be studied in general by analyzing
(unnormalized) factorial moments of the distribution $P_n$:
\[ 
  F_q = \sum_{n=q}^{\infty} n(n-1)\dots(n-q+1) P_n  ,         \label{facmom}
\]
or (unnormalized) factorial cumulant moments:
\[ 
  K_q = F_q - \sum_{i=1}^{q-1} {q-1 \choose i-1} K_{q-i} F_i .  \label{faccum}
\]
It was recently proposed to study their ratio\cite{DreminHwa2}:
\[ 
  H_q = K_q / F_q   .                               \label{hmom}
\]
which is indeed very sensitive to the properties 
of the tail of the distribution.
The theoretical basis for this analysis lies in  recent 
results\cite{DreminHwa2} obtained
in perturbative QCD, within the Modified Leading Log Approximation (MLLA),
which suggest that the ratio $H_q$ of eq.~\eref{hmom}
oscillates in sign when regarded as a function of the rank $q$.
On the contrary, in the case of the NBD, the ratio $H_q$ is always positive
and monotonically decreasing (see eq.~\eref{nbdhmom} below).
It was pointed out in ref.\ \cite{DreminHwa2} that many problems
appear  when one attempts to match the above mentioned QCD prediction
with experimental findings; the most prominent ones
are the inadequacy of MLLA calculations of moments of rank higher than
the second one (higher order
correlations are not under control in perturbative QCD), 
and the lack of knowledge of hadronization effects. 
Comparisons with experimental data have nonetheless
appeared in the recent past\cite{Gianini}, showing qualitative agreement 
with the
theoretical work. This analysis however suffers from the experimental
point of view of one more problem that does not apply to theory
\cite{Levchenko}: finite statistics does not allow a precise knowledge
of the shape of the high multiplicity tail, because one obtains a truncated MD.
This fact makes the straightforward experimental analysis questionable, 
as it will be shown in detail in the following section.

\section{Truncating the multiplicity distribution}
Let us consider a truncated multiplicity distribution $P_n$, which
is zero for $n > n_0$, $n_0$ being the cutoff parameter. It is clear
from the definition, eq.~\eref{facmom}, that $F_q = 0$ for $q > n_0$.
It is also clear from the definition, eq.~\eref{faccum}, that
$K_q$ is also in this case different from zero for all $q$. 
One can prove the following theorem about distributions and 
infinitely divisible distributions (the latter are, in our view, very important
in multiparticle dynamics\cite{void}):

\vspace{0.4\baselineskip}
\noindent {\em Theorem 1\/}: a truncated multiplicity distribution 
is not infinitely divisible.\\
The generating function of a discrete infinitely divisible distribution 
can always be cast in the exponential form
\begin{eqnarray}
f(z) &=& \E^{\lambda [g(z) - 1]} \nonumber \\
     &=& \E^{-\lambda} \sum_{N=0}^\infty \frac{1}{N!} \left[\lambda g(z)\right]^N
      = \E^{-\lambda} \sum_{N=0}^\infty \frac{\lambda^N}{N!} 
          \left[ \sum_{m=0}^\infty q_m z^m \right]^N
\end{eqnarray}
where $\lambda$ is a constant, positive parameter and $g(z)$ the generating
function for the probabilities $q_m$, so that all coefficients in the
summation are non-negative. Because $g(z)$ is a generating function,
one can always find $m_0$ such that $q_{m_0} > 0$. Then the coefficient
of $z^{Nm_0}$ (to which $P_{Nm_0}$ is proportional) is always larger
than or equal to $\E^{-\lambda} \lambda^N q_{m_0}^N / {N!}$
for every $N > 0$, therefore the distribution $P_n$
cannot be truncated. Hence a MD which is truncated cannot be infinitely
divisible.

A second theorem can be proven on factorial cumulant moments' properties
in a truncated MD:

\vspace{0.4\baselineskip}
\noindent {\em Theorem 2\/}: the factorial cumulant moments of a 
truncated MD are not positive definite.\\
The combinants of a MD can be defined as follows\cite{Kauffman}:
\[
    C_q = \frac{P_q}{P_0} - \frac{1}{q}\sum_{i=1}^{q-1} (q-i)C_{q-i}
                                  \frac{P_i}{P_0}
\]
The combinants of all ranks of a MD are all non-negative
if and only if the distribution is infinitely divisible\cite{Combin},
and since it was shown in Theorem 1 that this is not the case for a
truncated distribution, it follows that at least one of the $C_q$ is
negative.
But given the relation between factorial cumulant moments and 
combinants\cite{Kauffman}:
\[
    K_q = \sum_{j=q}^{\infty} j(j-1) \dots (j-q+1) C_j
\]
this implies that the $K_q$ are not positive definite. Notice that
since for a truncated MD $K_q$ can be negative and $F_q$ cannot,
their ratio $H_q$ can also change sign.

In order to visually compare the effects of truncating the multiplicity
distribution on the $H_q$, one can take an exact MD, truncate it
and compute the resulting moments. For the phenomenological reasons
mentioned in the introduction, we used the NBD, whose 
form and moments are well known in terms of the two parameters
$\nbar$ and $k$ ($\nbar$ is the average number
of particles and $k$ is linked to the dispersion $D$ by
$k^{-1} + \nbar^{-1} = D^2/\nbar^2$):
\[ 
    P_n^{\mathrm{NBD}} = \frac{k(k+1)\dots(k+n-1)}{n!}
      \left(\frac{\nbar}{\nbar+k}\right)^n 
      \left(\frac{k}{\nbar+k}\right)^k  \label{pnbd}
\]
\[ 
    F_q = \nbar^q \frac{(k+1)(k+2)\dots(k+q-1)}{k^{q-1}}   \label{nbdfac}
\]
\[ 
    K_q = \nbar^q \frac{(q-1)!}{k^{q-1}}               \label{nbdfcum}
\]
\[ 
    H_q = \frac{(q-1)!}{(k+1)(k+2)\dots(k+q-1)}       \label{nbdhmom}
\]
Experimental data on charged particles MD in full phase space only 
present even multiplicities because of charge conservation
(``even-odd effect''); therefore the
truncated form that will be discussed is
\[ 
    P_n = \cases{ A P_n^{\mathrm{NBD}} &if ($n$ even) and ($n \leq n_0$)\cr
                 0  &otherwise\cr}                     \label{trunc}
\]
Here $n_0$ is the maximum observed multiplicity and $A$ is the normalization
parameter, so that $\sum_n P_n = 1$; the parameters $\nbar$ and $k$ of
the complete NBD have been chosen  to reproduce various experimental
distributions. 

\begin{figure}
 \begin{center}%
\mbox{\epsfig{file=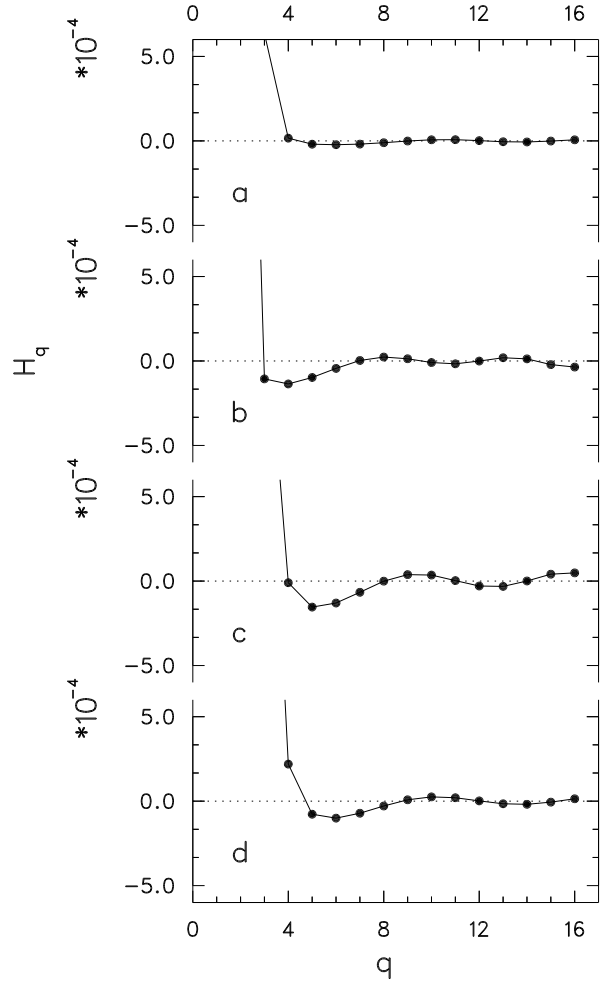,height=13cm,bbllx=99pt,bblly=85pt,bburx=425pt,bbury=565pt}}%
\end{center}
 \caption[noentry]{The ratio of factorial cumulant moments to factorial
moments $H_q$ is plotted as a function of the rank $q$
for the distribution in eq.~\protect\eref{trunc}, for different choices of
parameters, corresponding to a good NBD description of experimental data in
\ee\ annihilation, as listed in Table \protect\ref{tabee}.
The lines are drawn only to guide the eye. }\label{figee}
\end{figure}

In Figure \ref{figee}, $H_q$ is plotted as a function of $q$ for the
selection of values of the parameters
of the NBD appearing in eq.~\eref{trunc} 
listed in Table \ref{tabee}:
the values used were taken in order to reproduce the MD's in \ee\
annihilation experiments at different energies, and $n_0$ was 
taken as the largest
multiplicity in the published data for the same energy.
Various amplitudes of oscillation are seen, depending on the parameters.
Notice that the first minimum lies around $q = 5$ and
that the amplitude of the oscillations are comparable to those found
in the experimental data\cite{Gianini}.

\begin{table} 
 \begin{center}
 \caption[noentry]{Parameters used in Figure \protect\ref{figee}.
  The NBD with these values of $\nbar$ and $k$ reproduces well the
  MD of \ee\ annihilation
  data in different experiments and at different c.m.\ energies 
  as listed. $n_0$ is largest multiplicity in the published data.
   }\label{tabee}
 \vspace{4mm}
 \begin{tabular}{lrrrl}
  \hline
    & $\nbar$ & $k$  & $n_0$ &  ~Experiment \\ \hline
 a) & 12.9  & 212.0  & 28   & HRS (29 GeV) \cite{HRS:1} \\
 b) & 13.6  &  54.0  & 36   & Tasso (34 GeV) \cite{Tasso}\\
 c) & 15.5  &  30.8  & 38   & Tasso (43 GeV) \cite{Tasso}\\
 d) & 21.4  &  24.3  & 52   & Delphi (91 GeV) \cite{DEL:1} \\
  \hline
 \end{tabular}
 \end{center}
\end{table} 

\begin{figure}
 \begin{center}%
\mbox{\epsfig{file=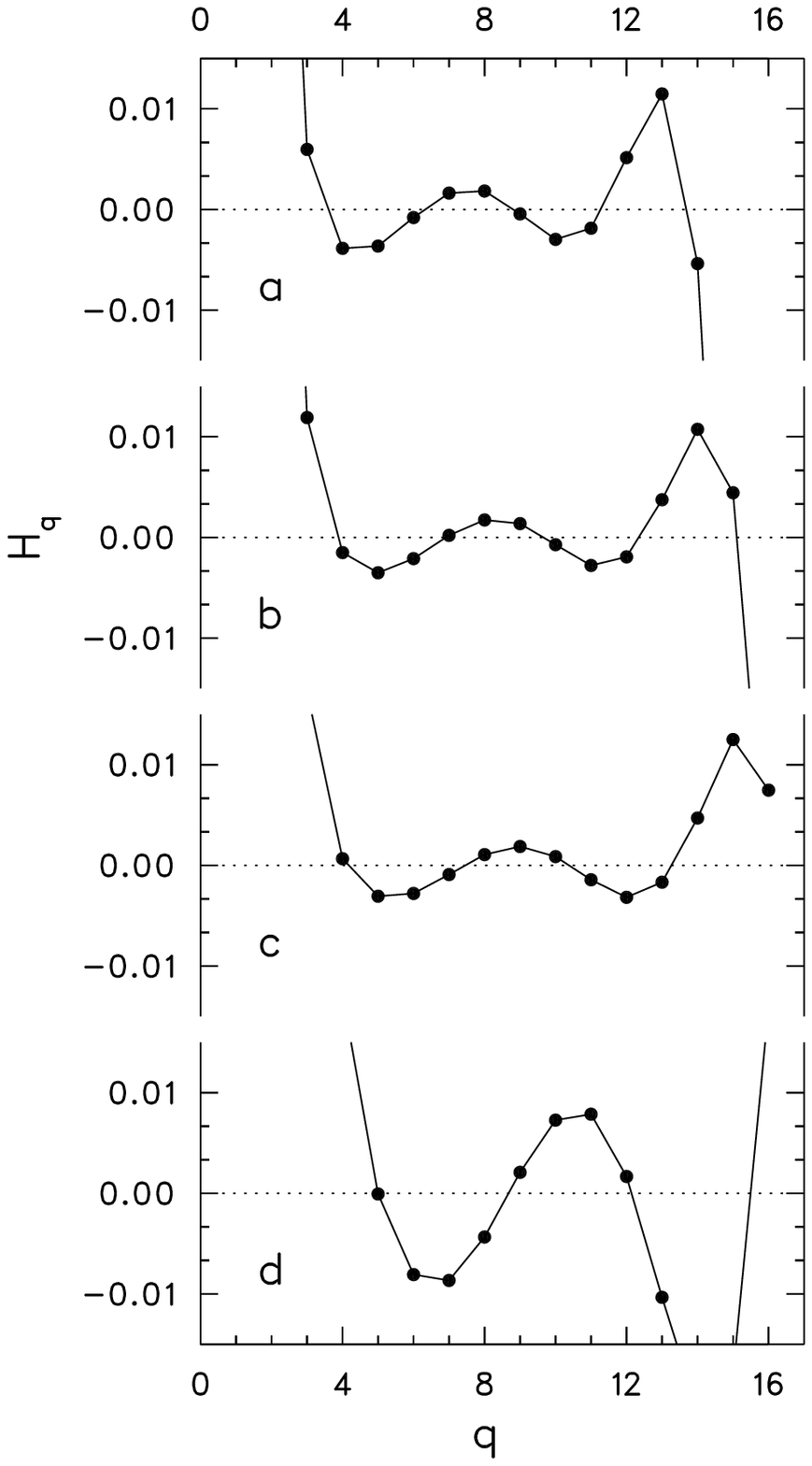,height=13cm,bbllx=99pt,bblly=85pt,bburx=425pt,bbury=565pt}}%
\end{center}
 \caption[noentry]{The ratio of factorial cumulant moments to factorial
moments $H_q$ is plotted as a function of the rank $q$
for the distribution in eq.~\protect\eref{trunc}, for different choices of
parameters, corresponding to a good NBD description of experimental data in
$pp$ and $\bar p p$ collisions, as listed in Table \protect\ref{tabpp}.
The lines are drawn only to guide the eye. }\label{figpp}
\end{figure}

\begin{table}
 \begin{center}
  \caption[noentry]{Parameters used in Figure \protect\ref{figpp}.
  The NBD with these values of $\nbar$ and $k$ reproduces well the
  MD of  $pp$ and $\bar pp$ collisions
  data in different experiments and at different c.m.\ energies 
  as listed. $n_0$ is largest multiplicity in the published data.
   }\label{tabpp}
 \vspace{4mm}
 \begin{tabular}{lrrrl}
  \hline
    & $\nbar$ & $k$  & $n_0$ &  ~Experiment \\ \hline
 a) & 10.7  &  11.0  & 26   & ISR (30 GeV) \cite{ISR:0} \\
 b) & 12.2  &   9.4  & 32   & ISR (44 GeV) \cite{ISR:0} \\
 c) & 13.6  &   8.2  & 38   & ISR (62 GeV) \cite{ISR:0} \\
 d) & 28.3  &   3.7  &100   & UA5 (540 GeV) \cite{UA5:rep} \\
  \hline
 \end{tabular}
 \end{center}
\end{table} 

Figure \ref{figpp} is similar to Figure \ref{figee}, but the parameters
are taken to reproduce MD's in $pp$ and $\bar pp$ collider events at different
energies (see Table \ref{tabpp}).
Oscillations here are larger than in the previous case
because a larger part of the MD is truncated away; 
their amplitudes and general shapes are
still comparable to their experimental counterpart\cite{Gianini}.

The unphysical effects of the finite statistics are clearly apparent
from these figures: this fact should be taken into consideration when
analyzing the ratio $H_q$ in experimental data, and these
effects should be removed from the data in order to expose physical
properties.

\begin{figure}
 \begin{center}%
\mbox{\epsfig{file=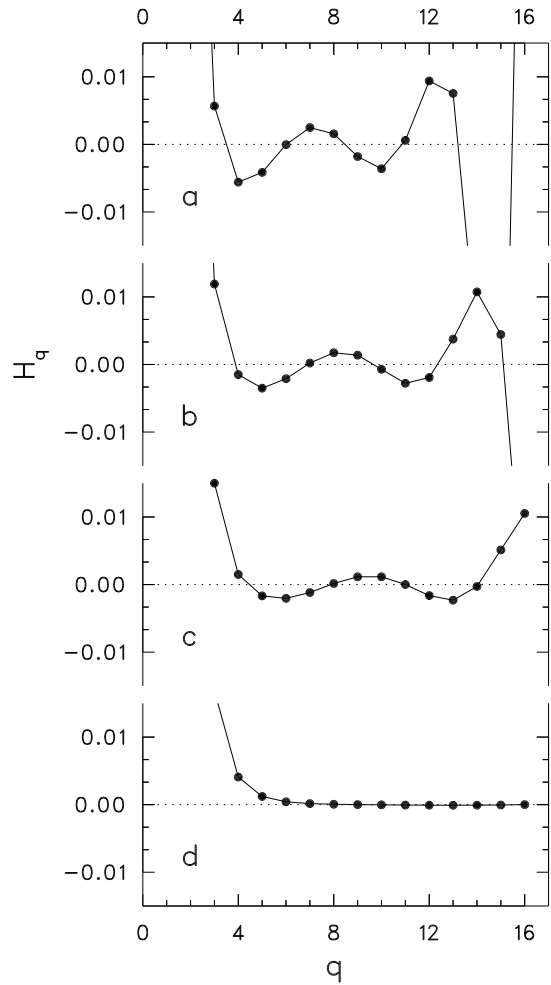,height=13cm,bbllx=99pt,bblly=85pt,bburx=425pt,bbury=565pt}}%
\end{center}
 \caption[noentry]{The ratio $H_q$ is plotted here for $\nbar = 12.2$,
 $k = 9.4$ and a) $n_0 = 28$, b) $n_0 = 32$, c) $n_0 = 36$, d) $n_0 = 64$.
 }\label{fign0}
\end{figure}

In Figure \ref{fign0} we show how the value of the cutoff $n_0$
affects the position and amplitude of the oscillations: the other
parameters are the same as in Figure \ref{figpp}b but $n_0$ grows
from 28 to 64 ($n_0 = 32$ in Figure \ref{figpp}b as in 
Figure \ref{fign0}b.) 
In the last case no oscillations are visible in the
figure, thus showing that such a large cutoff
does not affect appreciably the ratio $H_q$, but lower values,
comparable to the experimental maximum multiplicity, give larger
oscillations because they correspond to cutting away a larger part
of the distribution.

\section{Conclusions}
We have shown that the consequences of the natural fact that
the experimental multiplicity distribution is truncated because
of finite statistics have a large importance in the study of
the ratio of factorial cumulant moments to factorial moments, and
can mask  physical results.
An oscillatory behaviour of $H_q$ versus $q$ appears, which is comparable
in size and shape to that observed in a straightforward analysis of
experimental data as in ref.~\cite{Gianini}.
It is therefore suggested to take this effect into account 
in future analyses.

\newpage

%% References file: h.REF
%% creator: REFLATEX 1.0
%% Usage: \input this file where references should go

\end{document}